\catcode`\@=11
\expandafter\ifx\csname @iasmacros\endcsname\relax
	\global\let\@iasmacros=\par
\else	\immediate\write16{}
	\immediate\write16{Warning:}
	\immediate\write16{You have tried to input iasmacros more than once.}
	\immediate\write16{}
	\endinput
\fi
\catcode`\@=12


\def\rmb{\seventeenrm}

\def\singlespace{\baselineskip=\normalbaselineskip}
\def\halfspace{\baselineskip=1.5\normalbaselineskip}
\def\doublespace{\baselineskip=2\normalbaselineskip}


\def\AB{\bigskip\parindent=40pt
        \centerline{\bf ABSTRACT}\medskip\halfspace\narrower}
\def\AE{\bigskip\nonarrower\doublespace}
\def\nonarrower{\advance\leftskip by-\parindent
	\advance\rightskip by-\parindent}


\def\boxit#1{\vbox{\hrule\hbox{\vrule\kern3pt
	\vbox{\kern3pt#1\kern3pt}\kern3pt\vrule}\hrule}}

\def\hence{\leavevmode\hbox{\bf .\raise5.5pt\hbox{.}.} }

\def\dalemb#1#2{{\vbox{\hrule height.#2pt
	\hbox{\vrule width.#2pt height#1pt \kern#1pt \vrule width.#2pt}
	\hrule height.#2pt}}}
\def\gtorder{\mathrel{\raise.3ex\hbox{$>$}\mkern-14mu
             \lower0.6ex\hbox{$\sim$}}}
\def\ltorder{\mathrel{\raise.3ex\hbox{$<$}\mkern-14mu
             \lower0.6ex\hbox{$\sim$}}}

\newdimen\fullhsize
\newbox\leftcolumn
\def\twoup{\hoffset=-.5in \voffset=-.25in
  \hsize=4.75in \fullhsize=10in \vsize=6.9in
  \def\fullline{\hbox to\fullhsize}
  \let\lr=L
  \output={\if L\lr
        \global\setbox\leftcolumn=\columnbox\global\let\lr=R \advancepageno
      \else \doubleformat \global\let\lr=L\fi
    \ifnum\outputpenalty>-20000 \else\dosupereject\fi}
  \def\doubleformat{\shipout\vbox{
    \fullline{\box\leftcolumn\hfil\columnbox}\advancepageno}}
  \def\columnbox{\leftline{\vbox{\makeheadline\pagebody\makefootline}}}
  \tolerance=1000 }
\catcode`\@=11					



\font\fiverm=cmr5				
\font\fivemi=cmmi5				
\font\fivesy=cmsy5				
\font\fivebf=cmbx5				

\skewchar\fivemi='177
\skewchar\fivesy='60


\font\sixrm=cmr6				
\font\sixi=cmmi6				
\font\sixsy=cmsy6				
\font\sixbf=cmbx6				

\skewchar\sixi='177
\skewchar\sixsy='60


\font\sevenrm=cmr7				
\font\seveni=cmmi7				
\font\sevensy=cmsy7				
\font\sevenit=cmti7				
\font\sevenbf=cmbx7				

\skewchar\seveni='177
\skewchar\sevensy='60


\font\eightrm=cmr8				
\font\eighti=cmmi8				
\font\eightsy=cmsy8				
\font\eightit=cmti8				
\font\eightbf=cmbx8				

\skewchar\eighti='177
\skewchar\eightsy='60


\font\ninei=cmmi9
\font\ninesy=cmsy9

\skewchar\ninei='177
\skewchar\ninesy='60


\font\tenrm=cmr10				
\font\teni=cmmi10				
\font\tensy=cmsy10				
\font\tenex=cmex10				
\font\tenit=cmti10				
\font\tensl=cmsl10				
\font\tenbf=cmbx10				
\font\tentt=cmtt10				
\font\tenss=cmss10				
\font\tensc=cmcsc10				
\font\tenbi=cmmib10				

\skewchar\teni='177
\skewchar\tenbi='177
\skewchar\tensy='60

\def\tenpoint{\ifmmode\err@badsizechange\else
	\textfont0=\tenrm \scriptfont0=\sevenrm \scriptscriptfont0=\fiverm
	\textfont1=\teni  \scriptfont1=\seveni  \scriptscriptfont1=\fivemi
	\textfont2=\tensy \scriptfont2=\sevensy \scriptscriptfont2=\fivesy
	\textfont3=\tenex \scriptfont3=\tenex   \scriptscriptfont3=\tenex
	\textfont4=\tenit \scriptfont4=\sevenit \scriptscriptfont4=\sevenit
	\textfont5=\tensl
	\textfont6=\tenbf \scriptfont6=\sevenbf \scriptscriptfont6=\fivebf
	\textfont7=\tentt
	\textfont8=\tenbi \scriptfont8=\seveni  \scriptscriptfont8=\fivemi
	\def\rm{\tenrm\fam=0 }%
	\def\it{\tenit\fam=4 }%
	\def\sl{\tensl\fam=5 }%
	\def\bf{\tenbf\fam=6 }%
	\def\tt{\tentt\fam=7 }%
	\def\ss{\tenss}%
	\def\sc{\tensc}%
	\def\bmit{\fam=8 }%
	\rm\setparameters\setbaselines\fi}


\font\twelverm=cmr12				
\font\twelvei=cmmi12				
\font\twelvesy=cmsy10	scaled\magstep1		
\font\twelveex=cmex10	scaled\magstep1		
\font\twelveit=cmti12				
\font\twelvesl=cmsl12				
\font\twelvebf=cmbx12				
\font\twelvett=cmtt12				
\font\twelvess=cmss12				
\font\twelvesc=cmcsc10	scaled\magstep1		
\font\twelvebi=cmmib10	scaled\magstep1		

\skewchar\twelvei='177
\skewchar\twelvebi='177
\skewchar\twelvesy='60

\def\twelvepoint{\ifmmode\err@badsizechange\else
	\textfont0=\twelverm \scriptfont0=\eightrm \scriptscriptfont0=\sixrm
	\textfont1=\twelvei  \scriptfont1=\eighti  \scriptscriptfont1=\sixi
	\textfont2=\twelvesy \scriptfont2=\eightsy \scriptscriptfont2=\sixsy
	\textfont3=\twelveex \scriptfont3=\tenex   \scriptscriptfont3=\tenex
	\textfont4=\twelveit \scriptfont4=\eightit \scriptscriptfont4=\sevenit
	\textfont5=\twelvesl
	\textfont6=\twelvebf \scriptfont6=\eightbf \scriptscriptfont6=\sixbf
	\textfont7=\twelvett
	\textfont8=\twelvebi \scriptfont8=\eighti  \scriptscriptfont8=\sixi
	\def\rm{\twelverm\fam=0 }%
	\def\it{\twelveit\fam=4 }%
	\def\sl{\twelvesl\fam=5 }%
	\def\bf{\twelvebf\fam=6 }%
	\def\tt{\twelvett\fam=7 }%
	\def\ss{\twelvess}%
	\def\sc{\twelvesc}%
	\def\bmit{\fam=8 }%
	\rm\setparameters\setbaselines\fi}


\font\fourteenrm=cmr12	scaled\magstep1		
\font\fourteeni=cmmi12	scaled\magstep1		
\font\fourteensy=cmsy10	scaled\magstep2		
\font\fourteenex=cmex10	scaled\magstep2		
\font\fourteenit=cmti12	scaled\magstep1		
\font\fourteensl=cmsl12	scaled\magstep1		
\font\fourteenbf=cmbx12	scaled\magstep1		
\font\fourteentt=cmtt12	scaled\magstep1		
\font\fourteenss=cmss12	scaled\magstep1		
\font\fourteensc=cmcsc10 scaled\magstep2	
\font\fourteenbi=cmmib10 scaled\magstep2	

\skewchar\fourteeni='177
\skewchar\fourteenbi='177
\skewchar\fourteensy='60

\def\fourteenpoint{\ifmmode\err@badsizechange\else
	\textfont0=\fourteenrm \scriptfont0=\tenrm \scriptscriptfont0=\sevenrm
	\textfont1=\fourteeni  \scriptfont1=\teni  \scriptscriptfont1=\seveni
	\textfont2=\fourteensy \scriptfont2=\tensy \scriptscriptfont2=\sevensy
	\textfont3=\fourteenex \scriptfont3=\tenex \scriptscriptfont3=\tenex
	\textfont4=\fourteenit \scriptfont4=\tenit \scriptscriptfont4=\sevenit
	\textfont5=\fourteensl
	\textfont6=\fourteenbf \scriptfont6=\tenbf \scriptscriptfont6=\sevenbf
	\textfont7=\fourteentt
	\textfont8=\fourteenbi \scriptfont8=\tenbi \scriptscriptfont8=\seveni
	\def\rm{\fourteenrm\fam=0 }%
	\def\it{\fourteenit\fam=4 }%
	\def\sl{\fourteensl\fam=5 }%
	\def\bf{\fourteenbf\fam=6 }%
	\def\tt{\fourteentt\fam=7}%
	\def\ss{\fourteenss}%
	\def\sc{\fourteensc}%
	\def\bmit{\fam=8 }%
	\rm\setparameters\setbaselines\fi}


\font\seventeenrm=cmr10 scaled\magstep3		


\newdimen\rp@
\newcount\@basestretchnum
\newskip\@baseskip
\newskip\headskip
\newskip\footskip


\def\setparameters{\rp@=.1em
	\headskip=24\rp@
	\footskip=\headskip
	\delimitershortfall=5\rp@
	\nulldelimiterspace=1.2\rp@
	\scriptspace=0.5\rp@
	\abovedisplayskip=10\rp@ plus3\rp@ minus5\rp@
	\belowdisplayskip=10\rp@ plus3\rp@ minus5\rp@
	\abovedisplayshortskip=5\rp@ plus2\rp@ minus4\rp@
	\belowdisplayshortskip=10\rp@ plus3\rp@ minus5\rp@
	\normallineskip=\rp@
	\lineskip=\normallineskip
	\normallineskiplimit=0pt
	\lineskiplimit=\normallineskiplimit
	\jot=3\rp@
	\setbox0=\hbox{\the\textfont3 B}\p@renwd=\wd0
	\skip\footins=12\rp@ plus3\rp@ minus3\rp@
	\skip\topins=0pt plus0pt minus0pt}


\def\setbaselines{\maxdepth=4\rp@\baselinestretch=\@basestretchnum}


\def\baselinestretch{\afterassignment\@basestretch\@basestretchnum}
\def\@basestretch{%
	\@baseskip=12\rp@ \divide\@baseskip by1000
	\normalbaselineskip=\@basestretchnum\@baseskip
	\baselineskip=\normalbaselineskip
	\bigskipamount=\the\baselineskip
		plus.25\baselineskip minus.25\baselineskip
	\medskipamount=.5\baselineskip
		plus.125\baselineskip minus.125\baselineskip
	\smallskipamount=.25\baselineskip
		plus.0625\baselineskip minus.0625\baselineskip
	\setbox\strutbox=\hbox{\vrule height.708\baselineskip
		depth.292\baselineskip width0pt }}



\def\makeheadline{\vbox to0pt{\baselinestretch=1000
	\vskip-\headskip \vskip1.5pt
	\line{\vbox to\ht\strutbox{}\the\headline}\vss}\nointerlineskip}

\def\makefootline{\baselineskip=\footskip\line{\the\footline}}

\def\big#1{{\hbox{$\left#1\vbox to8.5\rp@ {}\right.\n@space$}}}
\def\Big#1{{\hbox{$\left#1\vbox to11.5\rp@ {}\right.\n@space$}}}
\def\bigg#1{{\hbox{$\left#1\vbox to14.5\rp@ {}\right.\n@space$}}}
\def\Bigg#1{{\hbox{$\left#1\vbox to17.5\rp@ {}\right.\n@space$}}}


\mathchardef\alpha="710B
\mathchardef\beta="710C
\mathchardef\gamma="710D
\mathchardef\delta="710E
\mathchardef\epsilon="710F
\mathchardef\zeta="7110
\mathchardef\eta="7111
\mathchardef\theta="7112
\mathchardef\iota="7113
\mathchardef\kappa="7114
\mathchardef\lambda="7115
\mathchardef\mu="7116
\mathchardef\nu="7117
\mathchardef\xi="7118
\mathchardef\pi="7119
\mathchardef\rho="711A
\mathchardef\sigma="711B
\mathchardef\tau="711C
\mathchardef\upsilon="711D
\mathchardef\phi="711E
\mathchardef\chi="711F
\mathchardef\psi="7120
\mathchardef\omega="7121
\mathchardef\varepsilon="7122
\mathchardef\vartheta="7123
\mathchardef\varpi="7124
\mathchardef\varrho="7125
\mathchardef\varsigma="7126
\mathchardef\varphi="7127
\mathchardef\imath="717B
\mathchardef\jmath="717C
\mathchardef\ell="7160
\mathchardef\wp="717D
\mathchardef\partial="7140
\mathchardef\flat="715B
\mathchardef\natural="715C
\mathchardef\sharp="715D


\def\err@badsizechange{%
	\immediate\write16{--> Size change not allowed in math mode, ignored}}

\baselinestretch=1000
\tenpoint

\catcode`\@=12					

\twelvepoint
\doublespace
{\nopagenumbers{
\rightline{IASSNS-HEP-97/96}
\rightline{~~~September, 1997}
\bigskip\bigskip
\centerline{\rmb Gauge Fixing in the Partition Function for}
\centerline{\rmb Generalized Quantum Dynamics}
\medskip
\centerline{\it Stephen L. Adler
}
\centerline{\bf Institute for Advanced Study}
\centerline{\bf Princeton, NJ 08540}
\medskip
\bigskip\bigskip
\leftline{\it Send correspondence to:}
\medskip
{\singlespace\leftline{Stephen L. Adler}
\leftline{Institute for Advanced Study}
\leftline{Olden Lane, Princeton, NJ 08540}
\leftline{Phone 609-734-8051; FAX 609-924-8399; email adler@ias.
edu}}
\bigskip\bigskip
}}
\vfill\eject
\pageno=2
\AB
We discuss the problem of gauge fixing for the partition function in 
generalized quantum (or trace) dynamics, deriving analogs of the 
De Witt-Faddeev-Popov procedure and of the BRST invariance familiar in the 
functional integral context.  
\AE
\bigskip\bigskip
\vfill\eject
\pageno=3
\centerline{\bf 1.~~Introduction, and the Gauge Theory} 
\centerline{\bf Axial Gauge Partition Function}

Generalized quantum (or trace) dynamics$^1$ is a generalization 
of classical mechanics, in which the Lagrangian and Hamiltonian are 
constructed as the trace of multinomials in noncommutative
operator or matrix variables.  This new form of analytic mechanics has a 
number of interesting properties.  
When the fermions are realized in the 
conventional manner as Grassmann matrices, generalized quantum dynamics 
incorporates and generalizes all rigid supersymmetry theories, 
such as$^2$ the Wess-Zumino and 
supersymmetric Yang-Mills theories, and also$^3$ the recently much discussed
``matrix model for M theory''.  In addition, one can formulate a 
statistical mechanics for generalized quantum dynamics, in both the 
canonical$^4$ and microcanonical$^5$ ensembles.  Recent work suggests$^{4,6}$ 
that 
this statistical mechanics can behave as a prequantum 
mechanics, with the Heisenberg commutation relations holding 
for statistical averages over the canonical ensemble 
of the underlying operator variables. 

This prior work on the statistical mechanics of generalized quantum 
dynamics has assumed either that one is dealing with an unconstrained 
system, or that the constraints have been explicitly integrated out.  
In the generic case for an unconstrained system, the canonical ensemble 
takes the form$^{4,5,6}$ 
$$\eqalign{
\rho=&Z^{-1}\exp(-\tau {\bf H}-{\rm Tr}\tilde \lambda \tilde C)~~~,\cr 
Z=&\int d\mu\exp(-\tau {\bf H}-{\rm Tr}\tilde \lambda \tilde C)~~~,\cr 
}\eqno(1a)$$
with $d\mu$ the invariant matrix (or operator) phase space 
measure provided$^4$ 
by Liouville's theorem, with ${\bf H}$ the conserved 
total trace Hamiltonian, and with 
$\tilde C$ the conserved operator 
$$\tilde C=\sum_{r ~~{\rm bosons}}[q_r,p_r] -\sum_{r~~{\rm fermions}}
\{q_r,p_r\}~~~.\eqno(1b)$$

In order to apply Eq.~(1a) directly to a constrained system, one must 
first explicitly integrate out the constraints.  A simple example where
this is possible, that forms the focus of this paper, is provided 
by the generalized quantum 
dynamics form of non-Abelian gauge theory,$^1$ which has a full operator 
valued 
gauge invariance group.  Let $A_{\nu}$ be an anti-self-adjoint operator 
or matrix valued gauge potential, let $F_{\mu\nu}$ be the corresponding 
gauge field strength defined by 
$$F_{\mu\nu}=\partial_{\mu}A_{\nu}-\partial_{\nu}A_{\mu}+[A_{\mu},A_{\nu}]
~~~,\eqno(2a)$$
and let ${\bf L}$ be the total trace Lagrangian defined by 
$${\bf L}={1\over 4} {\rm Tr}F_{\mu \nu}F^{\mu\nu}~~~.\eqno(2b)$$
As is well known, Eqs.~(2a, b) define a constrained dynamical system.
Introducing the canonical momenta 
$$p_{A_{\ell}}=-F_{0\ell}~,~~~\ell=1,2,3~~~,\eqno(3a)$$
and the covariant derivative 
$$D_{\mu}X=\partial_{\mu}X+[A_{\mu},X]~~~,\eqno(3b)$$
the constraint equation for the system of Eqs.~(2a, b) takes the form 
$$\sum_{\ell=1}^3 D_{\ell}p_{A_{\ell}}=0~~~.\eqno(3c)$$
Since in axial gauge, where $A_3=0$, the covariant 
derivative $D_3$ simplifies to $D_3=\partial_3$,   
the constraint of Eq.~(3c) is readily integrated out, giving as the explicit 
expression for the total trace Hamiltonian, 
$$\eqalign{
{\bf H}=&{\rm Tr}\left[-{1\over 2}\sum_{\ell=1}^3 p_{A_{\ell}}^2
-{1\over 4}\sum_{\ell,m =1}^3 F_{\ell m}^2\right] \cr
=&\int d^3x{1\over 2} {\rm Tr}\left[-\sum_{\ell=1}^2
p_{A_{\ell}}^2-F_{03}^2-F_{12}^2-\sum_{\ell=1}^2(\partial_3A_{\ell})^2 
\right]~~~,\cr
}\eqno(4a)$$
with $F_{03}$ given by the line integral
$$F_{03}=\int dz^{\prime}{1\over 2}{z-z^{\prime}\over|z-z^{\prime}|}
(\sum_{\ell=1}^2D_{\ell}p_{A_{\ell}})|_{z^{\prime}}~~~.\eqno(4b)$$
The conserved operator $\tilde C$ takes the form, in a general gauge, 
$$\tilde C=\int d^3x \sum_{\ell=1}^3 [A_{\ell},p_{A_{\ell}}]
=\sum_{\ell=1}^3 \left( \int d^3x D_{\ell}p_{A_{\ell}}
-\int dS_{\ell}p_{A_{\ell}} \right)~~~,\eqno(5a)$$
with $dS_{\ell}$ the surface element of the sphere at infinity.  To  
specialize Eq.~(5a) to axial gauge, one sets $A_3$ equal to zero and 
evaluates $p_{A_3}$  using Eqs.~(3a) and (4b).  
The axial gauge partition function is then given by 
Eq.~(1a), with the phase space measure $d \mu$ given by 
$$d \mu_{\rm axial}=\prod_{\vec x} \prod_{\ell=1}^2 dA_{\ell}(\vec x) 
dp_{A_{\ell}(\vec x)}~~~.\eqno(5b)$$

The problem we wish to address in this paper is how to generalize the 
axial gauge partition function to other gauges in which it may not be 
possible to explicitly integrate out the constraint.  This is a familiar 
problem in the theory of path integrals, and we shall use methods 
similar to the ones employed there to give a solution.  However, since 
the partition function singles out a Lorentz frame, we will have to make 
a restriction not encountered in the Lorentz scalar path 
integral case, namely we will consider only nontemporal gauge conditions that 
do 
not involve the scalar potential $A_0$.  This will still allow us to 
consider gauge transformations that rotate the axial gauge axis, or that 
transform to rotationally invariant gauges such as Coulomb gauge.  
We will also make  
the further assumption that the allowed gauge transformations 
leave invariant the surface integral contribution to ${\rm Tr} \tilde \lambda 
\tilde C$
arising from the second term on the right hand side of Eq.~(5a),    
which places a restriction on the constant of integration that governs 
the gauge transformation at the point at infinity.  
[Although the volume integral in the first term 
on the right hand side of Eq.~(5a) 
is generically gauge covariant rather than gauge invariant, 
this causes no problem 
because its integrand is proportional to the constraint of Eq.~(3c).]   
In Sec. 2 we develop an analog of the standard DeWitt-Faddeev-Popov 
method to write the axial gauge partition function in a general nontemporal 
gauge, subject to the surface term restriction just stated.  Since 
we have shown$^{2,3}$ that trace dynamics incorporates rigid supersymmetry, 
and since  BRST invariance is a particular rigid supersymmetry 
transformation, it is not surprising that the generalized expression for the 
partition function, when reexpressed in terms of ghost fermions,  
admits a BRST invariance, and this is demonstrated 
in Sec. 3.  In the analysis that follows, we do not attempt to address the 
issue of convergence of the partition function, which may well require 
significant restrictions on the class of trace Hamiltonians being considered.

\bigskip
\centerline{\bf 2. ~~General Nontemporal Gauges and }
\centerline{\bf the De Witt-Faddeev-Popov Procedure}
\bigskip
To express the partition function in a general nontemporal  
gauge, we follow closely the treatment of the De Witt-Faddeev-Popov 
construction in the familiar functional integral case, as given in 
the text of Weinberg$^7$.   Let us consider the integral
$$\eqalign{
Z_G=&\int d\mu B[f(A_{\ell})] \delta(D)\det{\cal F}[A_{\ell}]
\exp(-\tau {\bf H}-{\rm Tr}\tilde \lambda \tilde C)~~~,\cr
d\mu=& \prod_{\vec x} \prod_{\ell=1}^3 dA_{\ell}(\vec x)
dp_{A_{\ell}(\vec x)}~~~,\cr
}\eqno(6a)$$
with ${\bf H}$ given by the first line of Eq.~(4a) [which is valid in 
a general gauge on the constraint surface $D=0$ selected by the delta function 
in Eq.~(6a)], with $\tilde C$ given by Eq.~(5a), and with    
$D$ defined by  $D\equiv \sum_{\ell=1}^3D_{\ell}p_{A_{\ell}}$. 
The delta function of the anti-self-adjoint matrix valued argument $D$ 
appearing in Eq.~(6a) is given, in terms of 
ordinary delta functions of the real ($R$) and imaginary ($I$) parts of 
the matrix elements, by  
$$\delta(D)=\prod_{m<n} \delta((D_R)_{mn})\prod_{ m \leq n} 
\delta((D_I)_{mn})~~~,\eqno(6b)$$
and the integration measure over the anti-self-adjoint matrix 
$A_{\ell}$ is defined by 
$$dA_{\ell}=\prod_{m<n} d(A_{\ell R})_{mn}
\prod_{m \leq n} d(A_{\ell I})_{mn}~~~,\eqno(6c)$$
and similarly for $dp_{A_{\ell}}$.  The function $B[f]$ is an arbitrary 
integrable scalar valued function of the matrix valued 
argument $f(A_{\ell})$, 
which is used to specify the gauge condition.  We shall treat $f$ as a  
column vector $f_{\alpha}$ with $\alpha$ a composite index formed from 
the matrix row and column indices $m,n$; the argument ${\cal F}$ 
of the De Witt-Faddeev-Popov determinant is then given in terms of $f$ by 
the expression  
$${\cal F}_{\alpha \, \vec x,\beta \, \vec y}[A_{\ell}] 
\equiv { \delta f_{\alpha}(A_{\ell}(\vec x)+D_{\ell} \Lambda(\vec x))
\over \delta \Lambda_{\beta}(\vec y)}\vert_{\Lambda=0}~~~,\eqno(6d)$$
where $\delta$ is the usual functional derivative and 
$\beta$ is the composite of the row and column indices of the 
infinitesimal gauge transformation matrix $\Lambda$.  

We now demonstrate two properties of the integral $Z_G$ defined in Eq.~(6a):
(i) first, we show that when the gauge fixing functions $B[f]$ and 
$f(A_{\ell})$ are chosen to correspond to the axial gauge condition, 
then Eq.~(6a) reduces 
(up to an overall constant) to the axial gauge partition function of Sec.~1; 
(ii)  second, we show that $Z_G$ is in fact independent of the 
function $f(A_{\ell})$, and depends on the function $B[f]$ only through 
an overall constant.  These two properties together imply that 
$Z_G$ gives the wanted extension of the axial gauge partition function to 
general nontemporal gauges.  

To establish property (i), we make the conventional  axial gauge choice 
$$B[f(A_{\ell})]=\delta(A_3)=\prod_{m<n} \delta((A_{3R})_{mn})
\prod_{m \leq n}\delta((A_{3I})_{mn})
~~~,\eqno(7a)$$
so that 
$$\int dA_3 B[f(A_{\ell})]=\int dA_3 \delta(A_3)=1~~~.\eqno(7b)$$
With this gauge choice, 
$$D_3p_{A_3}=\partial_3p_{A_3}~~~,\eqno(7c)$$
which implies that 
$$\eqalign{
\delta(D)=&\delta(\partial_3p_{A_3}+\sum_{\ell=1}^2 D_{\ell}p_{A_{\ell}}) \cr
=&|\partial_3|^{-1}\delta(p_{A_3}+\int^z dz^{\prime} 
\sum_{\ell=1}^2 D_{\ell}p_{A_{\ell}} )~~~;\cr 
}\eqno(8a)$$
hence the integral over $p_{A_3}$ in $Z_G$ can be done explicitly, giving 
(up to an overall constant factor coming from the Jacobian $|\partial_3|^
{-1}$) the expression 
$$Z_G=\int d \mu_{\rm axial}  
\exp(-\tau {\bf H}-{\rm Tr}\tilde \lambda \tilde C)|_{A_3=0;~p_{A_3}=-\int^z  
dz^{\prime} \sum_{\ell=1}^2  D_{\ell}p_{A_{\ell}} }
~~~,\eqno(8b)$$
which agrees with the axial gauge partition function of Sec.~1.  

To establish property (ii), we first examine the gauge transformation 
properties of the various factors in the integral defining $Z_G$.  
We begin with the integration measure $d \mu$.  Under the infinitesimal gauge 
transformation (with $\Lambda$ anti-self-adjoint)
$$A_{\ell} \to A_{\ell}+D_{\ell}\Lambda=A_{\ell}+\partial_{\ell}\Lambda 
+[A_{\ell},\Lambda]~~~,\eqno(9a)$$ 
the inhomogeneous term $\partial_{\ell} \Lambda$ does not contribute 
to the transformation of the differential $dA_{\ell}$, and so 
$dA_{\ell}$ obeys the homogeneous transformation law $dA_{\ell} 
\to dA_{\ell}+\Delta_{\ell}$, with 
$$\Delta_{\ell}\equiv [dA_{\ell},\Lambda]~~~.\eqno(9b)$$
Hence to first order in $\Lambda$, the Jacobian of the transformation is 
$$\eqalign{
J=&1
+\sum_{m<n}{\partial(\Delta_{\ell R})_{mn} \over 
\partial (dA_{\ell R})_{mn} }  
+\sum_{m\leq n}{\partial(\Delta_{\ell I})_{mn} \over 
\partial (dA_{\ell I})_{mn} }  \cr
=&1
+(\sum_{m<n}+\sum_{m\leq n}) [(\Lambda_R)_{nn}-(\Lambda_R)_{mm}] \cr
=&1 ~~~,\cr
}\eqno(9c)$$
since the anti-self-adjointness of $\Lambda$ implies that $(\Lambda_R)_{nm}
=-(\Lambda_R)_{mn}$, and so the diagonal matrix elements 
$(\Lambda_R)_{nn}$ are all zero.  
Thus each factor $dA_{\ell}(\vec x)$ in the integration measure is 
gauge invariant. A similar argument applies to each factor 
$dp_{A_{\ell}(\vec x)}$ in the integration measure, and also to the factor 
$\delta(D)$ in the integrand,  since $D$ obeys the homogeneous gauge 
transformation law 
$D \to D +[D, \Lambda]$.  Turning to the exponential, the terms 
${\rm Tr} p^2_{A_{\ell}}$ and ${\rm Tr}F^2_{\ell m}$ are gauge invariant, 
and so the trace Hamiltonian ${\bf H}$ is gauge invariant.  From 
Eq.~(5a), the volume integral term in $\tilde C$ vanishes by virtue of 
the integrand factor $\delta(D)$, and so ${\rm Tr}\tilde \lambda \tilde C$ 
receives a contribution only from the surface term in Eq.~(5a), which by
hypothesis is left invariant by the class of gauge transformations under 
consideration.  In sum, we see that the integral $Z_G$ has the form 
$$Z_G=\int d\mu {\cal G}[A_{\ell}] B[f(A_{\ell})]\det {\cal F}[A_{\ell}]
~~~,\eqno(10a)$$ 
with the integration measure $d\mu$ and the integrand factor 
$${\cal G}[A_{\ell}]=\delta(D)\exp(-\tau{\bf H}-{\rm Tr} \tilde \lambda 
\tilde C)~~~\eqno(10b)$$
both gauge invariant.  Hence $Z_G$ has exactly the form assumed in the 
discussion of  Weinberg$^7$, and the proof given there 
completes the demonstration of property (ii).  
\vfill
\eject
\bigskip
\centerline{\bf 3.~~ Ghosts and BRST Invariance of the 
Generalized Partition Function}
\bigskip
Let us continue to follow the standard path integral analysis, and 
represent the De Witt-Faddeev-Popov determinant $\det {\cal F}[A_{\ell}]$
as an integral over fermionic ghosts, by writing 
$$\det {\cal F}[A_{\ell}]=\int d\omega^*d\omega
\exp(\int d^3x d^3y \omega^*_{\alpha}(\vec x)
{\cal F}_{\alpha \, \vec x,\beta \, \vec y}[A_{\ell}] 
\omega_{\beta}(\vec y))~~~,\eqno(11a)$$
with
$$d\omega=\prod_{\vec x}\prod_{m,n}d\omega_{mn}(\vec x)~~~,~~~ 
  d\omega^*=\prod_{\vec x}\prod_{m,n}d\omega^*_{mn}(\vec x)~~~. 
  \eqno(11b)$$
Let us also take for $B[f]$ the usual Gaussian 
$$B[f]=\exp(-{1 \over 2 \xi} \int d^3x {\rm Tr} f(A_{\ell}(\vec x))^2)
~~~,\eqno(11c)$$
and for $f(A_{\ell})$ the linear gauge condition 
$$f(A_{\ell})=\sum_{\ell} L^{\ell} A_{\ell}~~~,\eqno(11d)$$
in which $L^{\ell}$ can be either a fixed vector (such as $\delta_{\ell3}$ 
in axial gauge) or a differential operator (such as $\partial_{\ell}$ in 
Coulomb gauge), and a summation of $\ell$ from 1 to 3 is understood.     
With this choice of $f(A_{\ell})$, we find from Eq.~(6d)  
that 
$$\eqalign{
{\cal F}_{nm \vec x,pq \vec y}[A_{\ell}]=&
{\delta f_{nm}(A_{\ell}(\vec x)+D_{\ell}\Lambda(\vec x)) \over 
\delta \Lambda_{pq}(\vec y) } \cr
=&\sum_{\ell} L^{\ell}_{\vec x}\left({\partial \delta(\vec x-\vec y) \over 
\partial x^{\ell} } \delta_{np}\delta_{mq} 
+\delta(\vec x-\vec y)[(A_{\ell})_{np} \delta_{mq}
-\delta_{np}(A_{\ell})_{qm}] \right)~~~,\cr
}\eqno(12a)$$
which when substituted into the exponent in Eq.~(11a) gives, after  
integrations by parts, 
$$\int d^3x d^3y \omega^*_{\alpha}(\vec x)
{\cal F}_{\alpha \, \vec x,\beta \, \vec y}[A_{\ell}] 
\omega_{\beta}(\vec y))
=\int d^3x {\rm Tr} \overline{\omega}(\vec x) \sum_{\ell}L^{\ell}D_{\ell}
\omega(\vec x)~~~,\eqno(12b) $$
where we have defined $\overline {\omega}_{mn}=\omega^*_{nm}$.  
Hence the expression of Eq.~(10a) for $Z_G$ becomes 
$$Z_G=\int d\mu d\overline{\omega} d\omega {\cal G}[A_{\ell}] 
\exp[-\int d^3x{\rm Tr} ({1 \over 2 \xi} (\sum_{\ell}L^{\ell}A_{\ell})^2 -
\overline{\omega}(\vec x)\sum_{\ell}L^{\ell}D_{\ell}\omega(\vec x) )]
~~~~.\eqno(13a)$$
An alternative way of writing Eq.~(13a), that is convenient for exhibiting 
the BRST invariance, is to introduce an auxilliary self-adjoint matrix field 
$h$ and 
to reexpress Eq.~(13a) as 
$$Z_G=\int d\mu dh d\overline{\omega} d\omega {\cal G}[A_{\ell}] 
\exp[-\int d^3x{\rm Tr} ({\xi \over 2} h^2 + i h \sum_{\ell}L^{\ell}A_{\ell})
-\overline{\omega}(\vec x)\sum_{\ell}L^{\ell}D_{\ell}\omega(\vec x) )]
~~~~.\eqno(13b)$$

Starting from Eq.~(13b), we can now show that $Z_G$ has a BRST invariance 
of the familiar form.  Let $\theta$ be an $\vec x$-independent  
$c$-number Grassmann parameter 
(i.e., a $1 \times 1$ Grassmann matrix), and consider the variations 
defined by 
$$\eqalign{
\delta \omega=&\omega^2 \theta \cr
\delta A_{\ell}=&D_{\ell}\omega \theta \cr 
\delta \overline{\omega}=&-i h \theta \cr
\delta h =& 0 ~~~. \cr
}\eqno(14)$$
We begin by showing that Eq.~(14) defines a nilpotent transformation, 
in the sense that the second variations of all quantities are zero.  
To verify this, we show that the variations of $\omega^2$ and 
$D_{\ell}\omega$ are zero (the variations of $h$ and of $0$ are trivially  
0), as follows:
$$\eqalign{
\delta \omega^2=&\{ \delta \omega, \omega \}=\{\omega^2 \theta,\omega\}  
=\omega^2 \{\omega, \theta \}=0~~~, \cr
\delta D_{\ell}\omega=&[\delta A_{\ell},\omega] + D_{\ell} \delta \omega 
= [D_{\ell}\omega \theta, \omega] +D_{\ell} \omega^2 \theta 
=-\{D_{\ell} \omega,\omega\} \theta + \{D_{\ell}\omega ,\omega\} \theta =0
~~~.\cr
}\eqno(15a)$$
To see that $Z_G$ is invariant, we note that the action on $A_{\ell}$ 
of the BRST 
transformation of Eq.~(14) is just a gauge transformation 
(albeit with a Grassmann valued parameter), and so the gauge invariance 
analysis of Sec.~2 shows that the factors $d \mu$ and ${\cal G}[A_{\ell}]$ 
are  invariant.  The measure $dh$ is trivially invariant, and the measure 
$d \overline{\omega}$  is invariant because $\delta \overline{\omega}$ 
has no dependence on $\overline{\omega}$.  
Since 
$$\delta(d\omega)=d(\delta \omega)=d(\omega^2 \theta)=(\omega d\omega +
d\omega \omega)\theta~~~,\eqno(15b)$$
we have 
$$(\delta(d\omega))_{mn}=(\omega_{mm}d\omega_{mn}+d\omega_{mn} \omega_{nn})
\theta+...=d\omega_{mn}(\omega_{nn}-\omega_{mm})\theta+...,\eqno(15c) $$
with ... denoting terms that contain matrix elements $d\omega_{m^{\prime}
n^{\prime}}$ with  $(m^{\prime},n^{\prime}) \not= (m,n)$.  Hence the 
Jacobian of transformation for $d \omega$ 
differs from unity by a term proportional to 
$$\sum_{nm}(\omega_{nn}-\omega_{mm}) \theta =0~~~,\eqno(15d)$$
and so the measure $d\omega$ is also invariant.  Hence to complete the 
demonstration that $Z_G$ 
is BRST invariant, we have to show that the gauge fixing part of the 
Hamiltonian, 
$${\bf H}_G\equiv \int d^3x {\rm Tr}({\xi \over 2} h^2 
+i h \sum_{\ell}L^{\ell}A_{\ell}- \overline{\omega}\sum_{\ell}L^{\ell}
D_{\ell} \omega)~~~,\eqno(16a)$$
is BRST invariant.  Since we have already seen that $D_{\ell} \omega$
is invariant, and since $h$ is trivially invariant, we have only to verify 
that 
$$0=\int d^3x {\rm Tr}[ih\sum_{\ell}L^{\ell}\delta A_{\ell}
-(\delta \overline{\omega}) \sum_{\ell} L^{\ell} D_{\ell} \omega]
=\int d^3x {\rm Tr}ih \sum_{\ell}L^{\ell}D_{\ell}\{\omega, \theta\}~~~,
\eqno(16b)$$
which checks, completing the demonstration of BRST invariance of the  
generalized partition function.

\bigskip
\centerline{\bf Acknowledgments}
I wish to thank Lowell Brown for helpful conversations, and to acknowledge 
the hospitality of the Aspen Center for Physics, where this work was done. 
This work was supported in part by the Department of Energy under
Grant \#DE--FG02--90ER40542.
\vfill\eject
\centerline{\bf References}
\bigskip
\noindent
\item{[1]} S. L. Adler, Nucl. Phys.  {\bf B415}, 195 (1994);  
S. L. Adler, {\it Quaternionic Quantum Mechanics and 
Quantum Fields}, Secs. 13.5-13.7 (Oxford U. P., New York, 1995).
\bigskip 
\noindent
\item{[2]} S. L. Adler, preprint IASSNS-HEP-97/16, hep-th/9703132, 
Nucl. Phys. {\bf B} (in press).
\bigskip
\noindent
\item{[3]} S. L. Adler, preprint IASSNS-HEP-97/15, hep-th/9703053, 
Phys. Lett. {\bf B} (in press).
\bigskip
\noindent
\item{[4]}  S. L. Adler and A. C. Millard, Nucl Phys. {\bf B473}, 
199 (1996).
\bigskip
\noindent
\item{[5]} S. L. Adler and L. P. Horwitz, J. Math. Phys. {\bf 37}, 5429 
(1996).
\bigskip
\noindent
\item{[6]}  S. L. Adler and A. Kempf, preprint IASSNS-HEP-97/51, 
DAMTP-97-99, hep-th 9709106.
\bigskip
\noindent
\item{[7]}  S. Weinberg, {\it The Quantum Theory of Fields}, Vol. II, 
Sec. 15.5  (Cambridge U. P., Cambridge, 1996.)
\bigskip
\noindent
\bigskip
\noindent
\bigskip
\noindent
\bigskip
\noindent
\bigskip
\noindent
\bigskip
\noindent
\bigskip
\noindent
\bigskip
\noindent
\bigskip
\noindent
\bigskip
\noindent
\bigskip
\noindent
\bigskip
\noindent
\bigskip
\noindent
\bigskip
\noindent
\vfill
\eject
\bigskip
\bye